\documentclass{elsart}

\usepackage{graphicx,natbib,amssymb}
\journal{Advances in Space Research}
\begin{document}

\begin{frontmatter}


\title{The stable topology of the planetary systems of two 2:1 resonant companions:application to HD 82943}

\author[a,b]{JI  J.\corauthref{cor}},
\corauth[cor]{Corresponding author.}
\ead{jijh@pmo.ac.cn}
\author[c]{Kinoshita H.}, \author[d,b]{LIU L.}, \author[c]{Nakai H.}, \author[a,b]{LI G.}
\address[a]{Purple  Mountain Observatory, CAS, Nanjing 210008, China}
\address[b]{National Astronomical Observatory, CAS,  Beijing 100012, China}
\address[c]{National Astronomical Observatory, Mitaka, Tokyo 181-8588, Japan}
\address[d]{Department of Astronomy, Nanjing University, Nanjing 210093, China}

\begin{abstract}
We have numerically explored the stable planetary geometry for the
multiple systems involved in a 2:1 mean motion resonance, and
herein we mainly concentrate on the HD 82943 system by employing
two sets of the orbital parameters (Mayor et al. 2004; Ji et al.
2004). In the simulations, we find that all stable orbits are
related to the 2:1 resonance that can help to remain the
semi-major axes for two companions almost unaltered over the
secular evolution for $10^{8}$ yr. In addition, we also show that
there exist three possible stable configurations:(1) Type I, only
$\theta_{1} \approx 0^{\circ}$, (2) Type II,
$\theta_{1}\approx\theta_{2}\approx\theta_{3}\approx 0^{\circ}$
(aligned case), and (3) Type III, $\theta_{1}\approx 180^{\circ}$,
$\theta_{2}\approx0^{\circ}$, $\theta_{3}\approx180^{\circ}$
(antialigned case), here the lowest eccentricity-type mean motion
resonant arguments are $\theta_{1} = \lambda _{1} - 2\lambda _{2}
+ \varpi_{1}$ and $\theta_{2} = \lambda _{1} - 2\lambda _{2} +
\varpi_{2}$, the relative apsidal longitudes  $\theta_{3} =
\varpi_{1}-\varpi_{2}=\Delta\varpi$. And we find that other 2:1
resonant systems (e.g., GJ 876) may possess one of three stable
orbits in their realistic motions. Furthermore, we show that the
assumed terrestrial bodies cannot survive near the habitable zones
for HD 82943 due to strong perturbations induced by two resonant
companions, but such low-mass planets can be dynamically habitable
in the GJ 876 system at $\sim 1$ AU in the numerical surveys.
\end{abstract}
\end{frontmatter}

\section{Introduction}
\label{} The discovery of the extrasolar planets should show the
diversity of the planetary systems in the universe and also sketch
out an innovative world outside our solar system. Mayor \& Queloz
(1995) began such great revolution in our minds when they found
the first extrasolar giant Jupiter-51 Peg, and to date there are
more than 100 planetary systems (see also a daily updated
website\footnote{ http://cfa-www.harvard.edu/planets/bibli.html};
Butler et al. 2003), which most of them were detected with the
radial velocity technique in the surveys of nearby young stars. At
present, a dozen of multiple planetary systems were discovered in
recent years and this number is undoubtedly cumulative as time
elapses. Hence, it is necessary to categorize discovered multiple
systems according to their statistical characteristics (such  as
the distribution of the planetary masses, semi-major axes,
eccentricities and metallicity), then to study the correlation
between mass ratio and period ratio (Mazeh \& Zucker 2003) and to
improve  the understanding of the correlation between the planet
occurrence rate and stellar metallicity (Santos et al. 2003;
Fischer, Valenti, \& Marcy 2004). The other key point is to
investigate possible stable configurations for the multiple
systems, in which the observations reveal that most of them are
typically characterized by  mean motion resonance (MMR) and (or)
apsidal phase-locking between their  orbiting companions (Fischer
et al. 2003; Ji et al. 2003a), so  that one can better understand
the full dynamics of these systems.

\section{Stable Planetary Geometry in 2:1 Mean Motion Resonance}
Amongst the discovered extrasolar planetary systems, the resonant
pairs are quite common for the multiple systems, e.g., HD 82943
(Gozdziewski  \& Maciejewski 2001) and GJ 876 (Lee \& Peale 2002;
Ji, Li \& Liu 2002) in a 2:1 resonance, and 55 Cancri in a 3:1 MMR
(Marcy et al. 2002; Ji et al. 2003b). In addition, as the
observation results show that most of the systems can potentially
harbor more than one planet, thus we turn to the fascinating topic
of studying the stable planetary geometry locking into a
resonance, which such investigations are expected to be helpful in
an understanding of the orbital evolution of the planets and the
dynamical mechanisms of sustaining the stability of such systems.
In the present study, we mainly focus our attention on the HD
82943 system. In Table 1 are listed the orbital parameters for
this system, where two sets of the best-fit solutions are
respectively given according to Mayor et al. (2004) (Fit 2) and
their original orbital data (Fit 1; as of July 31, 2002), which
are adopted in our simulations.

In this paper, we employ N-body codes (Ji et al. 2002) to perform
direct numerical simulations by  using RKF7(8) and symplectic
integrators (Wisdom \& Holman 1991) for this system. In the
simulations, we always take the stellar mass and the minimum
planetary masses from Table 1. The time stepsize is usually
adopted to be $\sim$ 1\%-2.5\% of the orbital period of the
innermost planet, which is sufficiently smaller for the
integration. Additionally, the numerical errors were effectively
controlled with the accuracy of $10^{-14}$ over the integration
timescale.

\begin{table}{Table 1. Orbital parameters of HD 82943 planetary system.
The stellar mass $M_{c}$, for Fit 1, $M_{c}=1.05
M_{\odot}$, while for Fit 2, $M_{c}=1.15M_{\odot}$}
\begin{tabular}{lcllccc}
\hline \hline
Planet  &Mass(Mjup) &$P$(day)  &$a$(AU)  &$e$   &$\omega$(deg) &$\tau$(JD)\\
\hline
b (Fit 1) &1.63  &444.6(8.8)   &1.16    &0.41(0.08)  &96(7)   &2451620.3(12)\\
c (Fit 1) &0.88  &221.6(2.7)   &0.73    &0.54(0.05)  &138(13) &2451630.9(5.9)\\
\hline
b (Fit 2) &1.84  &435.1(1.4)   &1.18    &0.18(0.04)  &237(13) &2451758(13)\\
c (Fit 2) &1.85  &219.4(0.2)   &0.75    &0.38(0.01)  &124(3)  &2452284(1)\\
\hline \hline
\end{tabular}

\end{table}

In our simulations for HD 82943 system, we found three  possible
stable configurations for a system in a 2:1 MMR: (I) only
$\theta_{1}$ librates  about  $0^{\circ}$, (II)
$\theta_{1}\approx\theta_{2}\approx\theta_{3}\approx 0^{\circ}$
(aligned case), (III)  $\theta_{1}\approx 180^{\circ}$,
$\theta_{2}\approx0^{\circ}$, $\theta_{3}\approx180^{\circ}$
(antialigned case). The definition of the lowest eccentricity-type
2:1 resonant arguments are $\theta_{1} = \lambda _{1} - 2\lambda
_{2} + \varpi_{1}$ and $\theta_{2} = \lambda _{1} - 2\lambda _{2}
+ \varpi_{2}$. And the relative apsidal longitude is
$\theta_{3}=\Delta\varpi = \varpi_{1} - \varpi_{2}$, \noindent
where $\lambda _{1,2}$ are, respectively, the mean longitudes of
the inner and outer planets, and $\varpi_{1,2}$, respectively,
their periastron longitudes. Obviously, for above three angles, no
more than two are linearly independent. Besides, these stable
planetary geometry is related to the existence of symmetric stable
equilibrium solutions, where we find that other 2:1 resonant
systems (e.g., GJ 876) may possess one of the above three stable
orbits in their realistic motions.

As a paradigm, we simply show the dynamical evolution for one of
the types of the stable orbits for HD 82943 and the reader may
refer to Ji et al. (2004) for a detailed study. Figures 1 and 2
exhibit that the long-term orbital evolution for one set of stable
solutions derived from Fit 2, where $\theta_{1}$ and $\theta_{3}$
both librate about $0^{\circ}$ (Type II) for $10^{8}$ yr,
meanwhile the semi-major axes and eccentricities for two massive
planets perform slight vibrations. Still, one may understand that
the 2:1 resonance can help remain the semi-major axes for two
companions almost unaltered over the secular evolution (see Figs.
1 and 2), and the apsidal phase-locking between two orbits can
further enhance the stability for this system, because the
eccentricities are simultaneously preserved to restrain the
planets from frequent close encounters. Our results are well
consistent with previous works, e.g., Hadjidemetriou (2002)
numerically studied the families of periodic orbits for the HD
82943 system under the rotating framework, and indicated that two
kinds of families orbits can survive for the 2:1 resonant planets.
Moreover, Beauge et al. (2003) further showed the asymmetric
stable solutions for the 2:1 resonant system using both analytical
and numerical means. In summary, all the above outcomes imply the
potential planetary configurations that the extrasolar systems
involved in 2:1 resonance may hold.

In order to understand the dynamics of this system, in a
companying paper (see Ji et al. 2004), we analytically plot the
evolution of the eccentricities and the relative apsidal
longitudes. The semi-analytical contour chart exhibits a good
agreement with the numerical investigations of maintaining the
eccentricities with larger $\theta_{3}$-libration amplitudes (ref.
to Fig.4b of Ji et al. 2004; Type II), and the figure also implies
that the $\theta_{3}$-libration will be broken up if the libration
amplitudes exceed or approach the critical value of $90^{\circ}$
near the separatrix. Bois (2003, private communication) also
confirmed that the stable strip of the alignment for the initial
$\varpi_{1}-\varpi_{2}$ is much narrower than that of the
antialignment in the calculations for HD 82943.

\section{Habitable  Zones}
The Habitable zones (HZ) are generally convinced to be suitable
places for terrestrial planets that can provide the liquid-water,
subtle temperature and atmosphere environment, and other proper
conditions (Kasting et al. 1993), supporting the development and
biological evolution of life on their surfaces. And herein, we
performed numerical surveys to examine the potential habitable
zones for the systems (e.g. HD 82943 and GJ 876) with a stable
geometry in 2:1 resonance. We generated 100 seed planets that all
bear the same masses as Earth in each run. The distribution of the
initial orbital elements for these postulated planets are as
follows : they move on the much less inclined belt with the
relative inclination with regard to the orbital plane of the
resonant companions less than $5^{\circ}$, with 0.96 AU$\leq a
\leq 1.05$ AU and $0\leq e \leq0.1$.

Figure 3 summarizes the main results of the simulations for the HD
82943 system - no Earth-mass planet can survive  about 1 AU in the
final system, where 95\% of the orbits are usually ejected at $t
\le 10^{5}$ yr due to the gravitational scattering arising from
two close massive planets. The scenario of a typical ejected orbit
may replay that the semi-major axis $a$ grows from $\sim 1$ AU to
tens of AU, meanwhile the eccentricity $e$ undergoes a rapid
increase from $\sim 0.1$ to 1. Hence, the assumed Earth-like
planet is thrown out from the two-planet system in less than
$10^{5}$ yr. As for GJ 876, all tests about 1 AU are stable at
least for $10^6$ yr. Figure 4 exhibits the typical orbital
evolution for the Earth-like planet, both the semi-major axis and
the eccentricity execute small fluctuations about 1 AU and 0.06,
respectively, and the inclination also remains less than 2 degrees
in the same time span. And there is no sign to indicate that such
regular orbits at $\sim 1$ AU with low eccentricities will become
chaotic for much longer time even for the age of the star.

In addition, we also carried out extended integrations for the HD
82943 planetary system to investigate whether there may exist the
analogous structure of asteroidal belts that the residual
planetesimals are left in the system in later planetary formation.
We let several hundreds of test particles initially located at the
regions of the 2:1, 3:1 and 5:2 MMRs with the inner massive planet
and then performed the simulations for 1 Myr. The results show
that most of the small bodies can be removed at the dynamical time
$\sim 10^{4}$ yr, which suggests that no survivals can finally
stay at above three resonant regions. In this sense, it is quite
similar to Kirkwood gaps for the main belt asteroids in our solar
system. However, there is much prospect of the space observations
with high resolutions (e.g., Spitzer) revealing the evidence of
the disk debris structure in other extrasolar planetary systems in
future.

Acknowledgments   This work is financially supported by National
Natural Science Foundation of China (Grant No. 10203005, 10173006,
10233020) and Foundation of Minor Planets of Purple Mountain
Observatory.

\clearpage

\begin{figure}
  \centering {\includegraphics[bb=0 0 595 442, scale=0.7]{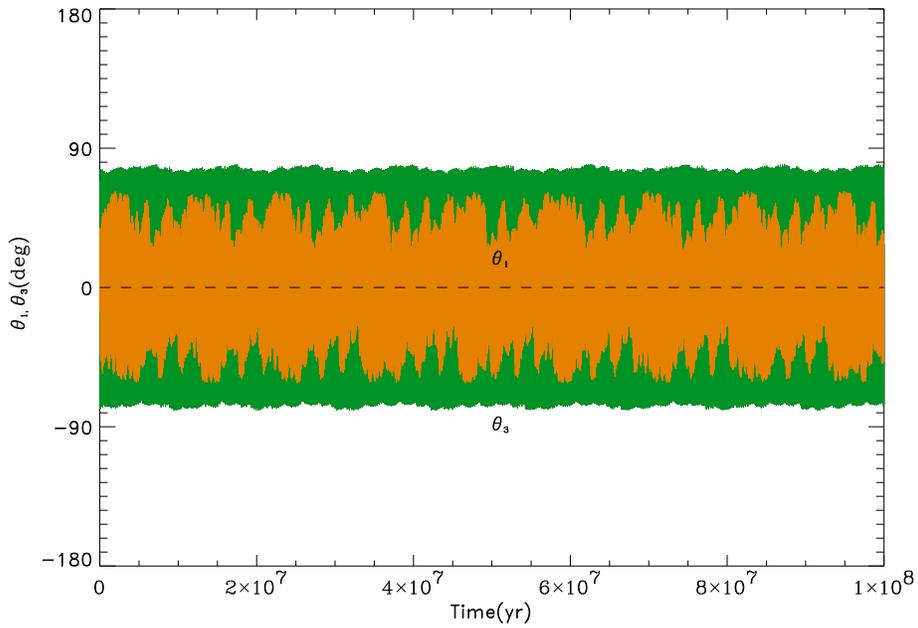}}
 \caption{One set of the stable solutions derived from Fit 2 (aligned orbits).
Long-term orbital evolution for $\theta_{1}$ and $\theta_{3}$,
where $\theta_{1}$ librates (by \textit{yellow color}) about
$0^{\circ}$ with a moderate amplitude of $\sim 60^{\circ}$,
$\theta_{3}$ (by \textit{green color}) about $0^{\circ}$ with a
larger amplitude of $\sim 70^{\circ}$ for $10^{8}$ yr, in
association with Type II orbit. The 2:1 resonance for this system
is confirmed by the modulations of the resonant angles, which
reveals the regular motion for two massive planets (see also
Figure 2).} \label{Figure 1}
\end{figure}

\begin{figure}
  \centering {\includegraphics[bb=0 0 595 442, scale=0.7]{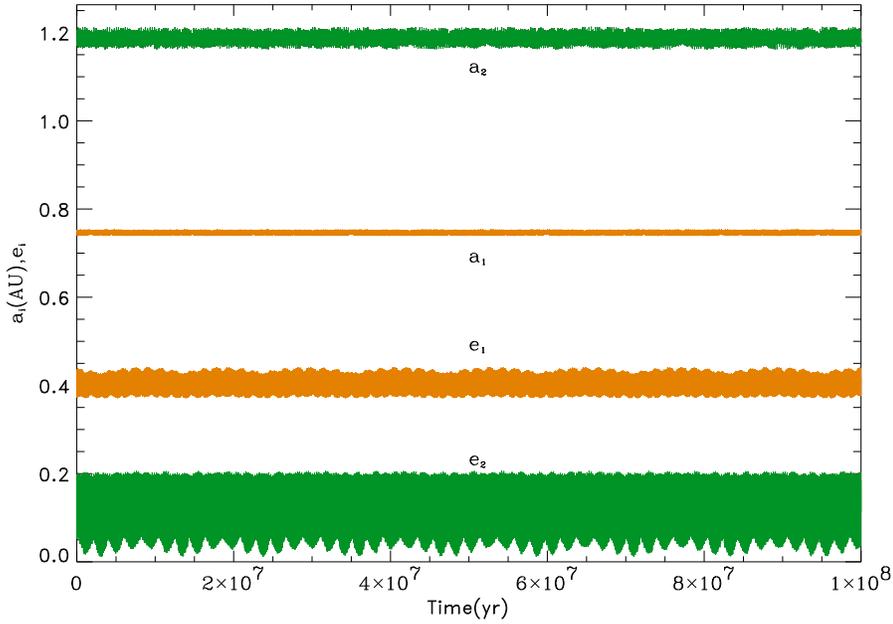}}
 \caption{Long-term orbital evolution for $a$ and $e$ for two planets
(the same as Figure 1, yellow and green colors, respectively, for
the inner and outer planets). The semi-major axes $a_{1}$ and
$a_{2}$ are almost unchanged and they modulate about 0.75 and 1.18
AU with relatively smaller amplitude for $10^{8}$ yr. In the same
time, the amplitude of the oscillations for $e_{1}$ and $e_{2}$
are not so large and they are just wandering in the span (0.35,
0.45) and (0, 0.20), respectively. No signs show that this system
will be chaotic for even longer evolution.} \label{Figure 2}
\end{figure}

\begin{figure}
  \centering {\includegraphics[bb=0 0 595 442, scale=1.0]{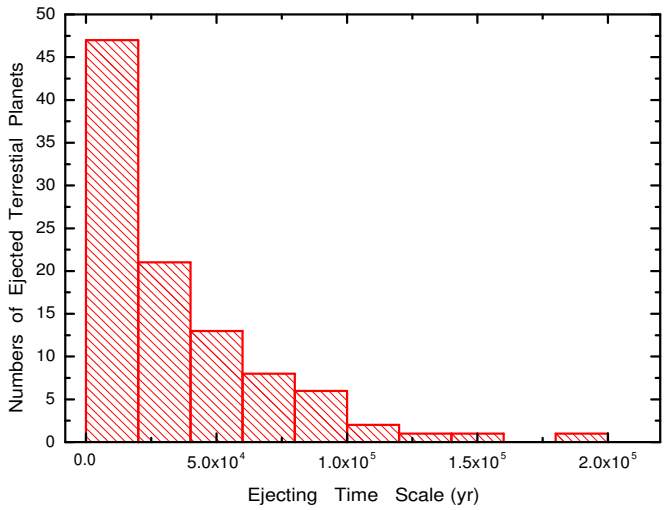}}
 \caption{The numbers of the ejected terrestrial planet vs ejecting
  timescale. Notice that 95\% of the
  orbits are ejected at $ t \le 1.0 \times 10^{5}$ yr.}
\label{Figure 4}
\end{figure}

\begin{figure}
  \centering {\includegraphics[bb=0 0 595 442, scale=0.7]{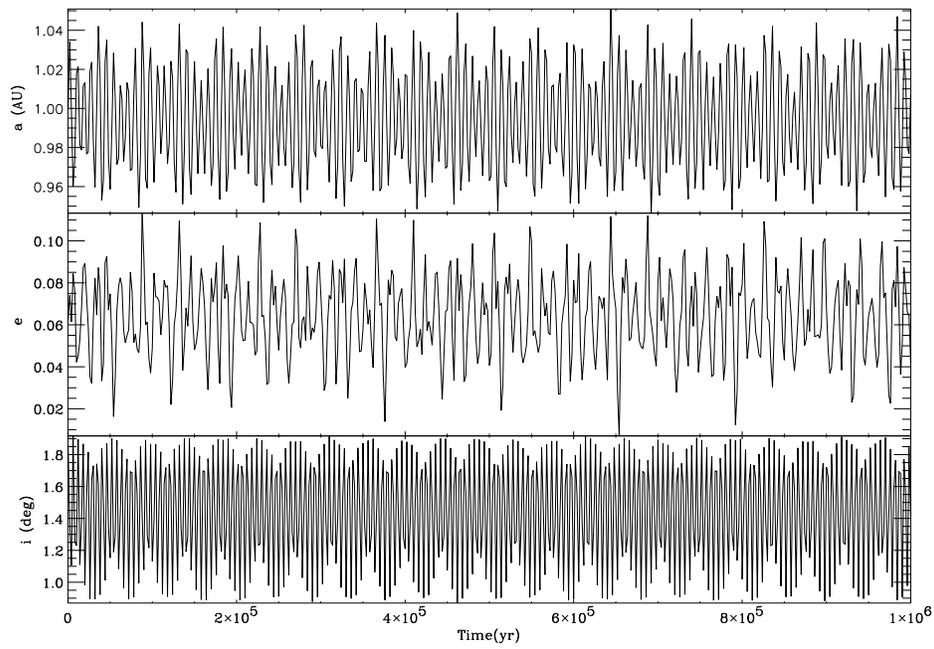}}
 \caption{ Orbital  evolution  of an Earth-mass  planet
placed in GJ 876 at 1 AU.  The  results  show that such orbits may
exist and be stable in the planetary system.} \label{Figure 5}
\end{figure}
\end{document}